\documentclass{article}
\usepackage{graphicx}
\usepackage{amsmath}

\begin{document}

\begin{center}
{\Large {\bf  Non-adiabatic Current Excitation in Quantum Rings }}
 \vskip5mm S. S. Gylfad\'{o}ttir$^{1}$\footnote{Email address: ssg@raunvis.hi.is},
V. Gudmundsson$^{1, 2}$, C. S. Tang$^{2}$, and A. Manolescu$^{1}$ \vskip3mm \mbox{}%
$^{1}$ Science Institute, University of Iceland,
       Dunhaga 3, IS-107 Reykjav\'{i}k, Iceland

\mbox{}$^{2}$ Physics Division, National Center for Theoretical Sciences,
P.O. Box 2-131, Hsinchu 30013, Taiwan

\bigskip

\begin{abstract}
We investigate the difference in the response of a
one-dimensional semiconductor quantum ring and a finite-width
ring to a strong and short-lived time-dependent perturbation in
the THz regime. In both cases the persistent current is modified
through a nonadiabatic change of the many-electron states of the
system, but by different mechanisms in each case.

\end{abstract}
\end{center}

\section*{1. Introduction}

Time-dependent or radiation induced phenomena in various
mesoscopic electron systems has attracted considerable interest
over the past few years~\cite{Ferry,Tang96,Tang99,Kom00,vg}.
These high-frequency modulated nanostructures, formed in a
two-dimensional electron gas (2DEG) by applying appropriate
confinement, are important both in connection with device
application and can function as convenient samples to probe 
the properties of electron systems in reduced dimensions. 

It is well discovered, in both theory~\cite{Buttiker83} and
experiment~\cite{Webb85}, that a quantum ring in a perpendicular
magnetic field or an Aharonov-Bohm magnetic flux can carry an
equilibrium current, the persistent current. This current is
periodic in the magnetic flux enclosed by the ring with a period
of the elementary flux quantum $\Phi_0=hc/e$. We present here a
method to change the persistent current pattern in quantum rings
non-adiabatically with a strong and short THz pulse. In this
paper we investigate a one-dimensional ring neglecting the
Coulomb interaction between spinless electrons, since its
inclusion in a LSDA scheme changes the results quantitatively,
but not qualitatively~\cite{vg,chakr}.

The results are compared to the results for a finite-width quantum
ring~\cite{vg} allowing us to identify different collective modes
in the excited state of the quantum rings influencing their
persistent currents.

\section*{2. Model and Approach}

We consider a one-dimensional ring with noninteracting spinless
electrons. The ring is subject to a uniform perpendicular
magnetic field ${\bf B} = B{\bf \hat{z}}$. At time $t=0$ the
system is in the ground state, the electron gas can be
described by the Hamiltonian
\begin{equation}
\begin{array}{rl}
    \displaystyle H_0&\displaystyle =\frac{1}{2m^*}\left[{\bf p}
    +\frac{e}{c}{\bf A}\right]^2\\
    &\\
    &\displaystyle = -\frac{\hbar^2}{2m^*r_0^2}\left[
    \frac{\partial^2}{\partial\theta^2}
    + 2i \ \frac{\Phi}{\Phi_0}\frac{\partial}{\partial\theta}
    - \left(\frac{\Phi}{\Phi_0}\right)^2\right]
\end{array}
\end{equation}
where $r_0$ is the radius of the ring, $\Phi=B\pi r_0^2$ is the
magnetic flux enclosed by it and $\Phi_0=hc/e$ is the unit flux
quantum. The eigenfunctions and the energy spectrum of the
Hamiltonian are given by
\begin{equation}
    \left<{\bf r}\,|\,l\,\right> = \psi_l(\theta)
    = \frac{e^{-il\theta}}{\sqrt{2\pi r_0}}\,,
    \qquad E_l = \frac{\hbar^2}{2m^*r_0^2}\left(l+\phi\right)^2,
\end{equation}
with $l=0, \pm1, \pm2,\dotsc$ and $\phi=\Phi/\Phi_0$. We use the
density matrix to specify the state of the system as the concepts
of a single electron energy spectrum and occupation number are
ill-defined when the system is subject to a strong external
time-dependent perturbation. The ground-state density matrix
$\rho_0$ is constructed from the expansion coefficients of
electronic states in terms of the basis states of $H_0$. At time
$t=0$ it is still possible to describe the occupation of the
states by the Fermi distribution and the chemical potential
$\mu$, ensuring the conservation of the number of electrons,
\begin{equation}
    \label{eq:densmat}
    \rho_{0,lm}=\sum_{\alpha}f(\epsilon_{\alpha}-\mu)\,c^*_{\alpha m}c_{\alpha l}
\end{equation}

At time $t=0$ the ring is radiated by a short THz pulse making the
Hamiltonian of the system time-dependent, namely $H(t)=H_0+V(t)$
with
\begin{equation}
    \label{eq:extpot1d}
    V(t)=V_0\cos\theta\,e^{-\Gamma t}
    \,\sin(\omega_1t)\,\sin(\omega t)\,\Theta(\pi-\omega_1t),
\end{equation}
where $\Theta$ is the Heaviside step function, and the parameters
are explained below. Here we consider the spatial part of the
excitation pulse as a dipole radiation. The equation of motion
for the density operator is given by
\begin{equation}
    i\hbar\frac{d}{dt}\rho(t)=[H+V(t),\rho(t)].
\end{equation}
The structure of this equation is inconvenient for numerical
evaluation so we resort instead to the time-evolution operator
$T$, defined by $\rho(t)=T(t)\rho_0T^{\dagger}(t)$, leading to a
simpler equation of motion
\begin{equation}
    i\hbar\,\dot{T}(t)=H(t)T(t)
\end{equation}
We then discretize the time variable and utilize the
Crank-Nicholson algorithm for the time integration~\cite{vg2} with
the initial condition, $T(0)=1$. This is performed in a truncated
basis of the Hamiltonian $H_0$, $\{\left|\,l\,\right>\}$. The
time-dependent orbital magnetization $M_0(t)$ is used to quantify
the currents induced by $V(t)$~\cite{Tan99:5626}.  The
magnetization operator ${\bf M}_0 = (1/2c){\bf r}\times{\bf j}$
together with the definition for the current, ${\bf j} =
-e\dot{\bf r} = (-ie/\hbar)[H(t),{\bf r}]$, give in terms of the
density matrix in the basis of $H_0$ the dynamic magnetization
\begin{equation}
\begin{aligned}
    &M_0(t)=-\frac{e}{2c}\text{tr}\{({\bf r}\times\dot{\bf r})\cdot\hat{\bf z}\,\rho(t)\}
    =-\frac{ie}{2\hbar c}\sum_{lmnp}\rho_{lm}(t)\\
    &\{\left<m\right|x\left|n\right>\left(\left<n\right|H(t)\left|p\right>
    \left<p\right|y\left|\,l\,\right>-\left<n\right|y\left|p\right>
    \left<p\right|H(t)\left|\,l\,\right>
    \right)\\
    &-\left<m\right|y\left|n\right>\left(\left<n\right|H(t)\left|p\right>
    \left<p\right|x\left|\,l\,\right>-\left<n\right|x\left|p\right>
    \left<p\right|H(t)\left|\,l\,\right>
    \right)\}
\end{aligned}
\end{equation}
which, on inserting for the matrix elements of $x=r_0\,\cos\theta$ and $y=r_0\,\sin\theta$,
for the 1D quantum ring, can be written
\begin{equation}
\begin{aligned}
    M_0(t)=\frac{er_0^2}{4\hbar c}\sum_{m,l}&\rho_{lm}\left(H_{m+1,l+1}(t)-H_{ml}(t)\right)\\
    &+\rho_{l,m+1}\left(H_{m+1,l}(t)-H_{m,l-1}(t)\right)
\end{aligned}
\end{equation}

The equations for the time-evolution operator are integrated for
a time interval of $4$ ps or longer using an increment of $0.0002$
ps and $M_0(t)$ is evaluated at each step. To model the ring with
a number of electrons, $N_e=1-8$, in a GaAs sample we select a
radius of $r_0=14$ nm and $m^*=0.067m$ where $m$ is the free
electron mass. For the radiation pulse of just over $3$ ps, we
select $\Gamma=2$ ps$^{-1}$ and the strength $V_0=0.29$ eV. The
envelope frequency corresponds to $\hbar\omega_1=0.658$ meV and
the base frequency is $\hbar\omega=2.63$ meV. We use a basis set
with $|\,l_{\rm max}|=50$ and the temperature is $T=4$ K.

\section*{3. Results and Discussion}

\begin{figure} [!hbp]
\includegraphics[clip=true,scale=0.5]{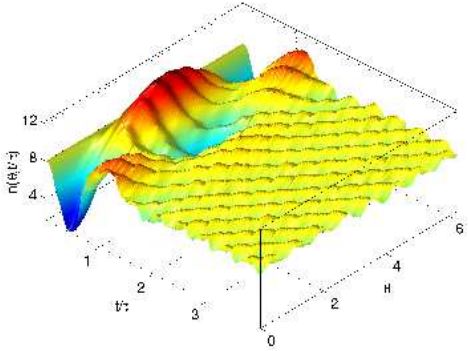}
\bigskip
\caption{Density as a function of the angle and time for 8 electrons on 
the ring in a magnetic field of $B=1.41$T.}
\label{Fig1}
\end{figure}

\begin{figure} [!hbp]
\includegraphics[clip=true,scale=0.5]{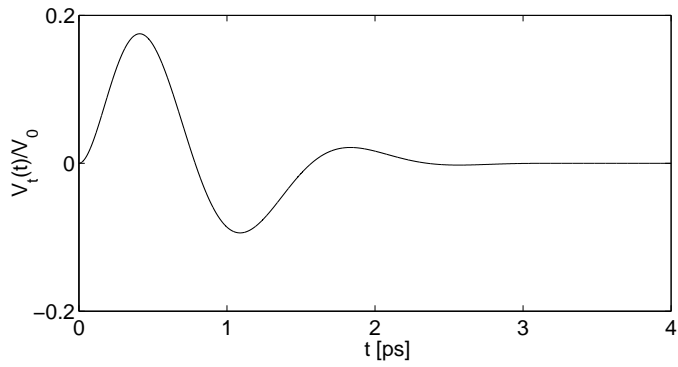}
\bigskip
\caption{The time-dependent part of the perturbation pulse. The space-dependent
part is a simple cosine.}
\label{Fig2}
\end{figure}

The time evolution of the electron density in the ring is shown
in Fig.\ \ref{Fig1} for a ring with eight electrons in a magnetic
field of $B=1.41$ T. We see clearly how the perturbation induces
angular oscillations that continue after the pulse has vanished.
The shape of the excitation pulse is shown in Fig.\
\ref{Fig2} in the time domain, and it should be remembered that
the spatial part of the pulse does not break the left/right
symmetry.

\begin{figure} [!hbp]
\begin{minipage}[t]{4.5cm}
\includegraphics[clip=true,scale=0.3]{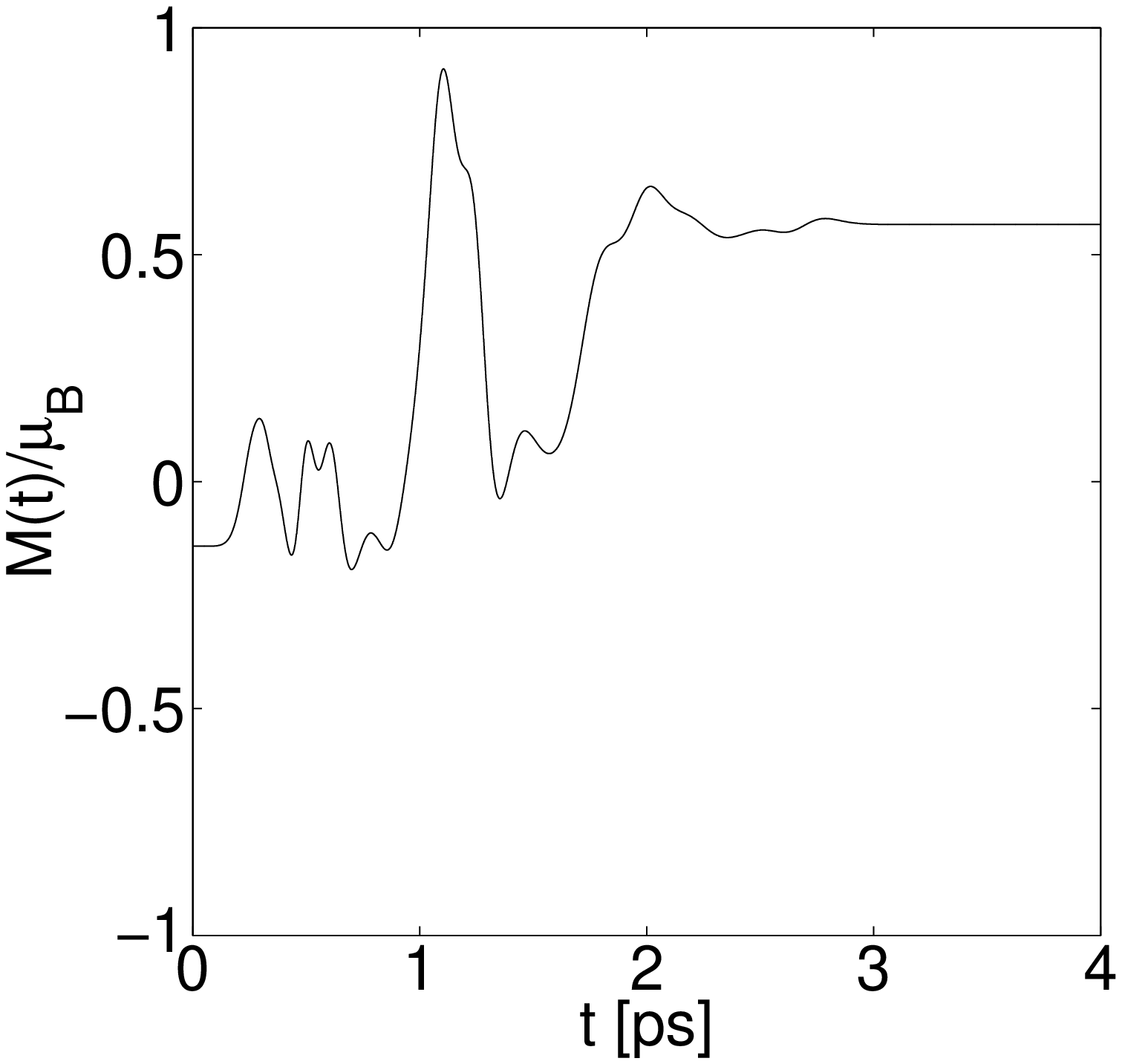}
\end{minipage}
\begin{minipage}[t]{4.5cm}
\includegraphics[clip=true,scale=0.3]{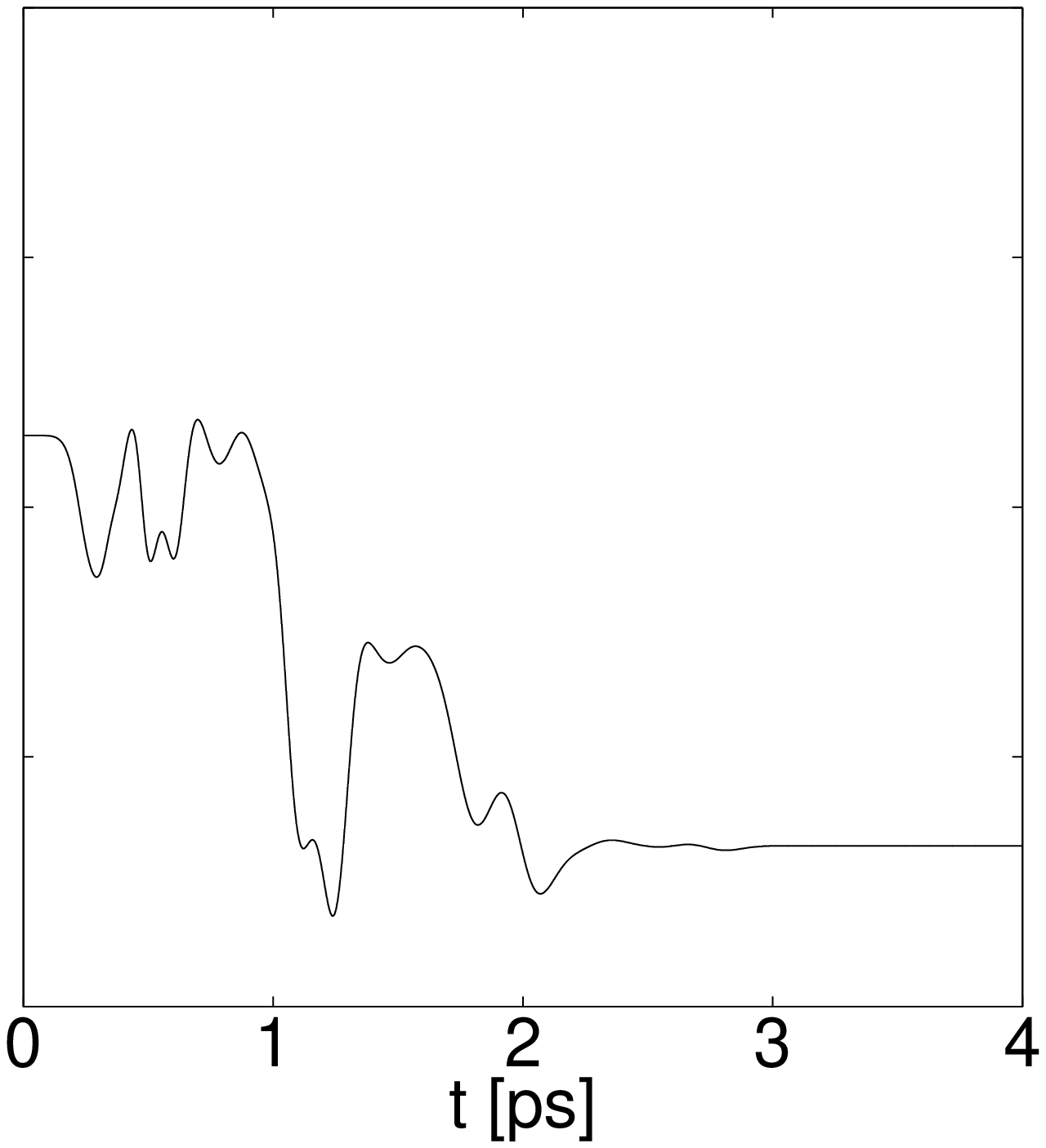}
\end{minipage}
\bigskip
\caption{Magnetization as a function of time for 1 electron and $B=1.71$T (left)
and $B=4.7$T (right).}
\label{Fig3}
\end{figure}

For the one-dimensional ring, the magnetization reaches a steady
state value almost immediately after the perturbation has been
turned off, as shown in Fig.\ \ref{Fig3}. Although there remain
oscillations in the electron density the magnetization remains
constant. This is different with the case of the finite-width ring
\cite{vg}, in which the radial plasma oscillations are excited by
the perturbation and the Lorentz force couples them with density
oscillations in the angular direction. As a consequence the
magnetization for the ring with finite width oscillates after the
excitation has been turned off, but its average value, the d.c.\
component is different from the initial value at $t=0$.

\begin{figure} [!hbp]
\includegraphics[clip=true,scale=0.5]{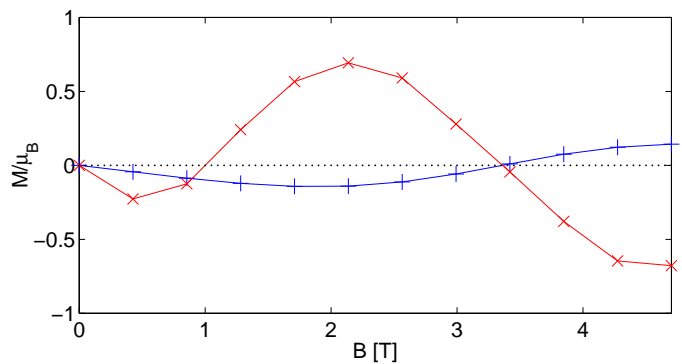}
\bigskip
\caption{Equilibrium magnetization $M_0(0)$ ($+$) and the magnetization
$M_0(t_f)$ ($\times$) after the radiation
pulse has vanished for one electron. Here $t_f$ denotes the final timestep.}
\label{Fig4}
\end{figure}

As for the finite-width ring the magnetization of the
one-dimensional ring after the perturbation has vanished is
different from the magnetization of the ground state for a
nonzero magnetic field, as shown in Fig.\ \ref{Fig4} for a single
electron ring. For low values of the magnetic field, the
magnetization (and thus the current) increases proportional to 
the perturbation strength. When $B\sim1$ T the magnetization changes sign,
implying that the direction of the current has changed, and for
all magnetic fields shown the current after excitation is in
anti-phase with the equilibrium persistent current.

\begin{figure} [!hbp]
\includegraphics[clip=true,scale=0.5]{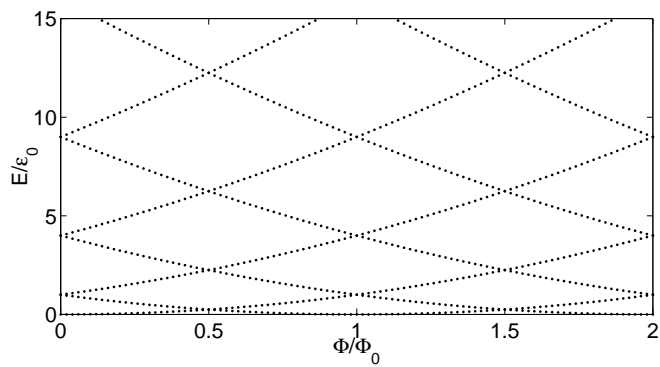}
\bigskip
\caption{Energy spectrum of the one-dimensional ring as a function of
the magnetic flux. The energy is given in units of $\epsilon_0=\hbar^2/2m^*r_0^2$}
\label{Fig5}
\end{figure}

At $B=3.36$ T (equivalent to $\phi=0.5$) the magnetization goes
to zero even after the system has been excited. The reason for
this becomes clear when looking at the energy spectrum of the
ring, shown in Fig.\ \ref{Fig5}. At integer and half-integer
values of the magnetic flux the slope of all energy levels is
zero. As the current is proportional to the slope, no states can
carry a current and thus it has no effect to excite the ring to a
higher energy state. At these values of $\phi$ the magnetic field
is not able to break the left/right symmetry of the system. The
spectrum of the finite-width ring is only quasi periodic in the
magnetic flux and thus the pulse is able to excite the system to
a higher energy state which has a finite magnetization at these
values of $\phi$.

\begin{figure}[!hbp]
\includegraphics[clip=true,scale=0.5]{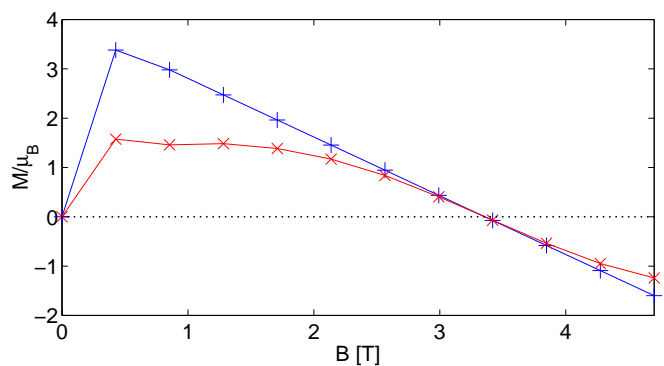}
\bigskip
\caption{Equilibrium magnetization $M_0(0)$ ($+$)
and the magnetization $M(t_f)$ ($\times$) after the radiation
pulse has vanished for 8 electrons.}
\label{Fig6}
\end{figure}

The current generated depends strongly on the number of electrons
on the ring. In Fig.\ \ref{Fig6}, we show the case of eight
electrons, the magnetization  decreases as a result of the
perturbation and it does not manage to alter the direction of the
current.

By inspecting the diagonal elements of the density matrix it
becomes clear that the persistent current is changed through a
modified occupation of the single-electron states, especially
around the chemical potential for the moderate excitation we use
here. After the perturbation has been turned off the combination
of angular momentum states are different from that of the ground
state.

\section*{4. Conclusions}

\noindent
The state of the system is changed nonadiabatically by the excitation
pulse. The persistent current of the system is changed through a modified
occupation of the single-electron states. After excitation it is
characteristic for a radiation induced
excited state with collective oscillations
and can thus be quite different from the equilibrium current.
The main difference between the 1D and the 2D quantum rings comes
from the different collective oscillation modes supported by the
systems. In the case of a ring with a finite
width the change in the persistent current is caused by
radial plasma oscillations coupled to the collective angular oscillations.
For both systems the breaking of the left/right symmetry that leads to a different
persistent current is not caused by the excitation pulse, but by the
external magnetic field.

\bigskip\noindent
{\large \bf Acknowledgements}

The research was partly funded by the Icelandic Natural Science Foundation,
the University of Iceland Research Fund, the National Science
Council of Taiwan under Grant No.\ 91-2119-M-007-004, and the
National Center for Theoretical Sciences, Tsing Hua University, Hsinchu Taiwan.
SSG would like to thank Dr. Ari Harju and Academy Professor Risto Nieminen
for their hospitality during a stay at the Laboratory of Physics of the
Helsinki University of Technology.

\end{document}